\documentclass[11pt,reqno]{amsart}
\usepackage{yhmath}
\usepackage{fancyhdr}
\usepackage{times}
\usepackage[T1]{fontenc}
\usepackage{mathrsfs}
\usepackage{latexsym}
\usepackage[dvips]{graphics}
\usepackage{epsfig}
\usepackage{hyperref, flafter}
\usepackage{amsmath,amsfonts,amsthm,amssymb,amscd}
\usepackage{color}
\usepackage{geometry}               
\usepackage{epstopdf}
\usepackage{verbatim}
\usepackage{bbm}

\newtheorem{thm}{Theorem}[section]

\newtheorem{lem}[thm]{Lemma}
\newtheorem*{thm*}{Theorem}
\newtheorem*{con*}{Conjecture}
\newtheorem*{lem*}{Lemma}
\newtheorem{prop}[thm]{Proposition}

\theoremstyle{definition}
\newtheorem{rmk}{Remark}

\def\L{\log T}
\def\t{\theta}
\def\s{\sigma}

\begin{document}
	
	\title[real moments log derivative] 	{Real moments of the logarithmic derivative of characteristic polynomials in random matrix ensembles}
	
	\author{Fan Ge}
	
	\address{Department of Mathematics, William \& Mary, Williamsburg, VA, United States}
	
	\email{ge@wm.edu}
	
	\keywords{Riemann zeta-function, logarithmic derivative, random matrices, real moments}
	
	\baselineskip=15pt
	
	\begin{abstract}
	We prove asymptotics for real moments of the logarithmic derivative of characteristic polynomials evaluated at $1-\frac{a}{N}$ in unitary, even orthogonal, and symplectic ensembles, where $a>0$ and $a=o(1)$ as the size $N$ of the matrix goes to infinity. Previously, such asymptotics were known only for integer moments (in the unitary ensemble by the work of Bailey, Bettin, Blower, Conrey, Prokhorov, Rubinstein and Snaith~\cite{bailey2019mixedmoments},  and in orthogonal and symplectic ensembles by the work of Alvarez and Snaith~\cite{alvarez2020moments}), except that in the odd orthogonal ensemble real moments asymptotics were obtained by Alvarez, Bousseyroux and Snaith~\cite{alvarez2023noninteger}. Our proof is new and does not make use of the known integer moments results in~\cite{bailey2019mixedmoments} and~\cite{alvarez2020moments}, and is different from the method in~\cite{alvarez2023noninteger} for the odd orthogonal ensemble.
	\end{abstract}
	
	\maketitle

\section{Introduction}

Zeros of the Riemann zeta-function have been of central importance in number theory. One way to investigate zeta zeros is to study the logarithmic derivative $\zeta'/\zeta$, which encodes information of zeta zeros via the Hadamard factorization formula. This brings considerable interest to the study of $\zeta'/\zeta$. Besides its relation to zeta zeros, the logarithmic derivative function $\zeta'/\zeta$ also offers connections between the zeros and the critical points of zeta; see for instance~\cite{levinson1974derivatives, soundararajan1998horizontal, zhang2001zeros, garaev2007small, ki2008zeros, farmer2012landau, radziwill2014gaps, ge2017number, ge2017gaps} in which the logarithmic derivative plays a key role in the study of critical points. In addition, $\zeta'/\zeta$ has intimate connections to the prime numbers. Indeed, Riemann's original plan for proving the Prime Number Theorem makes use of $\zeta'/\zeta$  in a crucial way. Futhermore, the mean values of $ \zeta'/\zeta$ are closely related to primes in short intervals; see~\cite{selberg1943prime, goldston2001pair}.

As is often the case, one seeks for simple quantities that can more or less represent important properties of the function itself, and in the theory of the Riemann zeta-function, mean values or moments often serve such purposes. To this end, let us define
	\begin{align*}
 \mathcal I_K(\sigma, T)=	\int_{T}^{2T} \left|\frac{\zeta'}{\zeta}\left(\sigma+it\right)\right|^{2K} dt .
\end{align*}
It turns out that the range of $\s$ has a crucial impact on the bahavior of $ \mathcal I_K(\sigma, T)$.  We separate the range of $\s$ into the following four cases. (Note that the average gap between zeta zeros near height $T$ is of order $1/\L$.)
\begin{itemize}
	\item The \textit{macroscopic} range: $  \s_0 \le \s \le 1$ for some constant $\s_0 >1/2$.  \\
	
	\item The \textit{mesoscopic} range: $\s = \dfrac{1}{2}+\dfrac{a}{\log T}$, where $a=o(\L) $ and $a\to \infty$ as $T\to\infty$. \\
	
	\item The \textit{microscopic} range: $\s = \dfrac{1}{2}+\dfrac{a}{\log T}$, where $a$ is of constant size. \\
	
	\item The \textit{nanoscopic} range: $\s = \dfrac{1}{2}+\dfrac{a}{\log T}$, where $a\to 0$ as $T\to\infty$. 
\end{itemize}

An important and well-known formula of Selberg~\cite{selberg1946contributions} expresses  $\zeta'/\zeta$ in terms of a prime sum plus a typically controllable error, provided that $\s$ is not too close to the critical line. In the macroscopic and mesoscopic range, it follows from Selberg's formula that $\zeta'/\zeta$ can be approximated by a short prime sum for most $t$. If we assume the Riemann Hypothesis (RH) then such approximation holds for all $t$, and as a result, the $2K$-th moment $\mathcal I_K$ of $\zeta'/\zeta$ can be approximated by the $2K$-th  moment of the prime sum, which can be computed for positive integers $K$ using the mean value theorem of Montgomery and Vaughan~\cite{mont1974hilbert}. Moreover,  in the mesoscopic range the distribution of $\zeta'/\zeta$  obeys  a central limit theorem; see  Lester~\cite{lester2014distribution} and Guo~\cite{guo1996zeros}.

It seems that in the macroscopic and mesoscopic range, the behavior of $\zeta'/\zeta$ is not much affected by the vertical distribution of zeta zeros, and thus it reflects less such information of them. The situation changes dramatically when we move closer to the critical line, that is, to the microscopic range. In this range, the behavior of $\zeta'/\zeta$ is largely affected by the distribution of zeta zeros. In fact, Goldston, Gonek, and Montgomery~\cite{goldston2001pair} showed that an asymptotic estimate of $\mathcal I_1$ in this range is \textit{equivalent} to Montgomery's Pair Correlation Conjecture. See also~\cite{selberg1943prime, goldston1987pair, goldston1988pair} for connections to primes in short intervals. Therefore, obtaining asymptotic behavior for moments of $\zeta'/\zeta$ in the microscopic range is highly important, but we also expect it to be extremely difficult. Indeed, it is also known that $\mathcal I_K$ has close connections to other important properties of zeta and primes (see Farmer, Gonek, Lee and Lester~\cite{farmer2013logderiv} for a connection to higher correlation functions of zeta zeros, Baluyot~\cite{baluyot2016ah} for a discussion of $\mathcal I_1$ under the Alternative Hypothesis, and Ki~\cite{ki2008zeros} for an estimate of $\mathcal I_1$ under an assumption on zeros of $\zeta'(s)$, as well as~\cite{soundararajan1998horizontal, zhang2001zeros, ge2017gaps} for implication of Ki's assumption to small gaps between zeta zeros), yet obtaining an unconditional asymptotic estimate for $\mathcal I_K$ in the microscopic range seems to be far out of reach.  In fact, even assuming knowledge on correlation functions of zeta zeros, it is still not clear how to obtain such asymptotic estimates for a general integer $K$; see~\cite{farmer2013logderiv} for some relevant discussion.

The extent to which the behavior of $\mathcal I_K$  reflects the distribution of typical sizes of zero gaps fades down when we move even closer to the critical line, namely, to the nanoscopic range. The first result illustrating such type of phenomena is a result of Goldston et al~\cite{goldston2001pair} who showed that an asymptotic for $\mathcal I_1$ is equivalent to the Essential Simplicity hypothesis, which roughly speaking asserts that there are not too many small gaps between zeta zeros. Therefore, the second moment of $\zeta'/\zeta$ is only capable of recognizing certain information of very small zero gaps, rather than their typical sizes. This result was generalized by the author~\cite{ge2023zeta} to all even integer moments, in which we established asymptotic estimates for $\mathcal I_K$ for all integers $K$ conditional on RH and a so-called $2K$-tuple Essential Simplicity hypothesis. This result also provides a conditional proof of a conjecture of Bailey, Bettin, Blower, Conrey, Prokhorov, Rubinstein and Snaith~\cite{bailey2019mixedmoments}. Thus, in the nanoscopic range, on one hand the quantities $\mathcal I_K$  capture only partial information on vertical gaps between zeros, but on the other hand the asymptotics for $\mathcal I_K$ become available under reasonable assumptions.

The above mentioned conjecture of Bailey et al was motivated by their investigation of analogous problems in the setting of random unitary matrices (also called the Circular Unitary Ensemble, or CUE for short). It is widely believed that the characteristic polynomials $P$ in the unitary ensemble $U(N)$ model the Riemann zeta-function regarding zero statistics, value distribution, and more (for example, see~\cite{montgomery1973pair} and~\cite{keating2000momentszeta}). In particular,  it is expected that the logarithmic derivative $P'/P$ of the characteristic polynomials models $ \zeta'/\zeta$ and is therefore  also important and interesting. What is good about CUE is that, unlike zeta, the distribution of zeros of $P$ is known, and thus, it is often hopeful to obtain unconditional results whose analogues are formidable in the zeta case.  Corresponding to the moments of $\zeta'/\zeta$, let us define for $a\in \mathbb R$
\begin{align*}
	\mathcal J_K = \mathcal J_K(a, N) =	\int_{U(N)} \left|\frac{P'}{P}\Big(1-\frac{a}{N}\Big)\right|^{2K} ,
\end{align*}
where the integral is taken with respect to the Haar measure on $U(N)$. Note that in the unitary ensemble the distribution of eigenvalues is invariant under rotation, and therefore, the integral remains unchanged if we introduce an extra averaging over the circle centered at the origin with radius $1-a/N$, which would look more directly like an analogue of $\mathcal I_K$ in the zeta case. The four ranges in this context are as follows.
\begin{itemize}
	\item The \textit{macroscopic} range: $0<\frac{a}{N}<1$ is of constant size.  \\[-2ex]
	
	\item The \textit{mesoscopic} range: $a=o(N)$ and $a \to \infty$ as $N\to\infty$.  \\[-2ex]
	
	\item The \textit{microscopic} range:  $a$ is of constant size.  \\[-2ex]
	
	\item The \textit{nanoscopic} range:  $a\to 0$ as $T\to\infty$. 
\end{itemize}

An important set of closed form formulas, known as the ratios theorems (see~\cite{conrey2008correlations, conrey2008autocorrelationofratios}), is useful for evaluating integrals like $\mathcal J_K$. Indeed, in the macroscopic and mesoscopic ranges the leading term of $\mathcal J_K$ for positive integers $K$ may be obtained by a direct application of the ratios formula. Moreover, like zeta, in the mesoscopic range we have a central limit theorem for $P'/P$ as proved in~\cite{ge2024cue}. 
The microsopic range is again the most difficult to deal with: while we have the powerful ratios theorems, deriving asymptotic estimates in the microscopic range for $\mathcal J_K$ from the ratios theorems is not easy. In fact, the situation becomes significantly more complicated as $K$ grows, and it is still an unsolved problem how to obtain asymptotics for $\mathcal J_K$ for a general $K\in\mathbb N$. See~\cite{bailey2019mixedmoments, farmer2013logderiv} for some relevant discussion. Now let us move closer to the unit circle and consider the nanoscopic range. In this range, although we may lose finer information of typical zero gaps (as suggested by the zeta case), it becomes possible to obtain unconditional asymptotics for $\mathcal J_K$. Indeed, in~\cite{bailey2019mixedmoments} Bailey et al proved that if $K$ is a positive integer and $a >0$ with $a\to 0$ as $N\to \infty$, then	
\begin{align}\label{thm bailey}
\mathcal J_K
	 \sim \dfrac{N^{2K}}{(2a)^{2K-1}} \cdot \binom{2K-2}{K-1}
\end{align}
as $N\to \infty$. Based on this, they also conjectured a similar estimate for $\mathcal I_K$ in the zeta case, and this is the aforementioned conjecture of theirs which was studied in~\cite{ge2023zeta}.

The conjectural connection between zeta and the unitary ensemble was developed further by Katz and Sarnak~\cite{katz1999random} into a more general philosophy that families of $L$-functions can be modelled by random matrices of certain symmetry types. This brings considerable interest to the study of orthogonal and symplectic ensembles as well. In~\cite{alvarez2020moments} Alvarez and Snaith proved analogous results of~\eqref{thm bailey} for orthogonal and symplectic random matrices. More precisely, for even orthogonal ensemble $SO(2N)$ their result is  
\begin{align}\label{thm a-s even o}
	\int_{SO(2N)} \left(\frac{P'}{P}\Big(1-\frac{a}{N}\Big)\right)^{K} 
	\sim (-1)^K \dfrac{2N^{K}}{a^{2K-1}} \cdot \dfrac{(2K-3)!!}{(K-1)!} 
\end{align}
for integers $K\ge 2$, 
while for odd orthogonal ensemble $SO(2N+1)$ they showed that
\begin{align}\label{thm a-s odd o}
	\int_{SO(2N+1)} \left(\frac{P'}{P}\Big(1-\frac{a}{N}\Big)\right)^{K} 
	= (-1)^K \left[ \left(\frac{N}{a}\right)^K - \frac{N^K}{a^{K-1}}K \right] + O\left(\frac{N^{K-1}}{a^{K-1}} + \frac{N^{K}}{a^{K-2}}\right) \quad 
\end{align}
for integers $K\ge 1$.
For symplectic ensemble $USp(2N)$ their result is 
\begin{align}\label{thm a-s s}
	\int_{USp(2N)} \left(\frac{P'}{P}\Big(1-\frac{a}{N}\Big)\right)^{K} 
	\sim (-1)^K \frac{2}{3} \dfrac{N^{K}}{a^{K-3}} \cdot \dfrac{(2K-5)!!}{(K-1)!} 
\end{align}
for integers $K\ge 4$.
We remark that in their work Alvarez and Snaith also obtained asymptotic results for $K=1$ in $SO(2N)$ and for $K=1,2,3$ in $USp(2N)$, but these results have different shapes from the above.

It is desirable to extend the above asymptotics to real moments (see the introduction of~\cite{alvarez2023noninteger} for some discussion on this matter). The method used in proving~\eqref{thm bailey}, \eqref{thm a-s even o}, \eqref{thm a-s odd o} and~\eqref{thm a-s s} in~\cite{bailey2019mixedmoments, alvarez2020moments} makes use of ratios theorems to transform the integral, and therefore is capable of treating integer moments but  does not generalize to real moments directly.
Very recently, Alvarez, Bousseyroux and Snaith~\cite{alvarez2023noninteger} extended the corresponding result for the odd orthogonal ensemble to real moments. Precisely, they proved that~\eqref{thm a-s odd o} is true for all real $K>0$. The reason that they can treat real moments for the odd orthogonal ensemble is that in $SO(2N+1)$ all matrices have a fixed eigenvalue at $1$. It turns out that this fixed eigenvalue at $1$ gives the main contribution in moments, and they are able to control the smaller order terms by making use of the integer moments result~\eqref{thm a-s odd o}. Note that in other ensembles we do not have any fixed eigenvalues, and as a result, the method in~\cite{alvarez2023noninteger} does not work directly for unitary, even orthogonal, or symplectic ensembles.

The purpose of this paper  is to prove real moments asymptotics for unitary, even orthogonal, and symplectic ensembles, as follows. 
	
\begin{thm}\label{thm u}
	 Let  $K> 1$ be a real number, and $a>0$ with $a\to 0$ as $N\to \infty$. Then	
\begin{align}\label{thm eq U}
	\notag				\int_{U(N)} \left|\frac{P'}{P}\Big(1-\frac{a}{N}\Big)\right|^{K} 
	& \sim \dfrac{N}{2\pi} \cdot \int_{-\infty}^{\infty} \left(\dfrac{1}{\left(\frac{a}{N}\right)^2 + \t^2} \right)^{\frac{K}{2}} d\t \\[1ex]
	& = \dfrac{N}{2\pi} \cdot \left(\dfrac{N}{a}\right)^{K-1} \cdot \sqrt{\pi} \cdot  \frac{ \Gamma \left(\frac{K-1}{2}\right)}{ \Gamma \left(\frac{K}{2}\right)}  \  .
\end{align}
\end{thm}

	\begin{thm}\label{thm so}
    	 Let  $K > 1$ be a real number, and $a>0$ with $a\to 0$ as $N\to \infty$. Then	
	\begin{align}\label{thm eq SO}
	\notag	\int_{SO(2N)} \left|\frac{P'}{P}\Big(1-\frac{a}{N}\Big)\right|^{K} 
		& \sim \dfrac{2N}{\pi} \cdot \int_{0}^{\infty} \left(\dfrac{2\cdot {\left(  \frac{a}{N}\right) }}{\left(\frac{a}{N}\right)^2 + \t^2} \right)^{K} d\t \\[1ex]
		& =   \dfrac{2N}{\pi} \cdot 2^K \cdot  \left(\dfrac{N}{a}\right)^{K-1} \cdot \frac{\sqrt{\pi}}{2} \cdot  \frac{ \Gamma \left(K-\frac{1}{2}\right)}{ \Gamma \left(K\right)}  \  .
	\end{align}
\end{thm}

\begin{thm}\label{thm usp}
	 Let  $K> 3$ be a real number, and $a>0$ with $a\to 0$ as $N\to \infty$. Then	
\begin{align}\label{thm eq USp}
\notag	\int_{USp(2N)} \left|\frac{P'}{P}\Big(1-\frac{a}{N}\Big)\right|^{K} 
	& \sim \dfrac{2N^3}{3\pi} \cdot \int_{0}^{\infty} \left(\dfrac{2\cdot {\left(  \frac{a}{N}\right) }}{\left(\frac{a}{N}\right)^2 + \t^2} \right)^{K} \cdot \t^2 d\t \\[1ex]
	& =   \dfrac{2N^3}{3\pi} \cdot 2^K \cdot  \left(\dfrac{N}{a}\right)^{K-3} \cdot \frac{\sqrt{\pi} }{4}\cdot  \frac{ \Gamma \left(K-\frac{3}{2}\right)}{ \Gamma \left(K\right)}  \  .
\end{align}
\end{thm}

\begin{rmk}
	In Theorems~\ref{thm so} and~\ref{thm usp} we consider $|P'/P|^K$ rather than $(P'/P)^K$ as in~\eqref{thm a-s even o} and~\eqref{thm a-s s}. 
For $K>1$ in the case of $SO(2N)$ and $K>3$ in the case of $USp(2N)$, the difference between $|P'/P|^K$ and $(P'/P)^K$ is minor. This is because from the proof of the theorems we can see that, for such $K$, the main contributing quantities in the integrals have the same sign. We choose to work with $|P'/P|^K$ because it avoids unnecessary complication caused by complex logarithm.
\end{rmk}

\begin{rmk}
Our proofs are new and do not make use of integer moments results~\eqref{thm bailey}, \eqref{thm a-s even o} or \eqref{thm a-s s}. By evaluating the gamma functions at half integers explicitly, one easily checks that our theorems agree with the known integer moments.
\end{rmk}

\begin{rmk}
Our method is robust for obtaining asymptotics for real moments when the parameter $K$ is not too small. The reason we require $K$ to be not too small is that, roughly speaking, the smaller $K$ is, the less influential large values of $P'/P$ are on the $K$-th moment. As our method eventually extracts the large values and uses them to compute moments, there is a threshold for $K$ above which we are able to do so. While this approach does not seem to work for $|P'/P|^K$ when $K$ is small, the results for $(P'/P)^K$ for small integers $K$ are known (see the remark following equation~\eqref{thm a-s s}). 
\end{rmk}

\begin{rmk}
	Some ideas in our proofs could be adapted to treat nanoscopic real moments in the zeta case, extending the integer moments results in~\cite{ge2023zeta}. The $2K$-tuple essential simplicity hypothesis may be replaced by a similar hypothesis for $2\lceil K \rceil$-tuples. Such hypotheses are used to control error terms in the context of zeta. We also note that the method in~\cite{ge2023zeta} for zeta may be adapted to treat integer moments in the case of the unitary ensemble, by introducing an additional average over the unit circle (and this is because the unitary ensemble is rotation-invariant). However, this approach does not work for orthogonal or symplectic ensembles. 
\end{rmk} 

We shall give a complete proof for the unitary case, and indicate necessary changes in other ensembles.

	\section{Unitary ensemble}
Recall that $a>0$ and $a=o(1)$ as $N\to \infty$. Let
	\begin{align*}
		z_0 = 1 - \frac{1}{N}  \quad \text{ and }  \quad  z = 1 - \frac{a}{N}.
	\end{align*}
We will need to choose a suitable parameter $c$ such that 
\begin{align}
	c&=o(1) \label{c 1}, \\
	a&=o(c), \label{c 2} \\
	\text{and } \quad c^{-K}&=o(a^{1-K}).  \label{c 3}
\end{align}
 For example, we may take
\begin{align*}
		c  =a^{\frac{K-1}{2K}}.
\end{align*}
Note that these restrictions of $c$ require $K>1$.

	Let $z_j = e^{i\t_j}$ be the eigenvalues and  write 
	\begin{align*}
		\frac{P'}{P}(z) = \sum_{|\t_j|<\frac{c}{N}} \frac{1}{z-z_j} + X_1 +X_2 -X_3,
	\end{align*}
	where
	\begin{align*}
		X_1 &= 	\frac{P'}{P}(z_0) = \sum_{j=1}^N \frac{1}{z_0-z_j}, \\
		X_2 &=  \sum_{|\t_j|\ge \frac{c}{N}} \left( \frac{1}{z-z_j} - \frac{1}{z_0-z_j} \right),\\
		X_3 &= \sum_{|\t_j|<\frac{c}{N}} \frac{1}{z_0-z_j}.
	\end{align*}
We require the following three lemmas. 
\begin{lem}\label{lem x1}
Let $\ell\in \mathbb Z^+$. Then 
\begin{align} \label{eq X1}
		\mathbb E |X_1|^{2\ell}   \ll_\ell N^{2\ell}. 
\end{align}
\end{lem}
This is Propostion~2.1 in~\cite{ge2024cue}.

\begin{lem}\label{lem x2}
We have 
\begin{align*}
	X_2 \ll \frac{1}{c} \cdot (N+|X_1|) 
\end{align*}
\end{lem}
\noindent \textit{Proof}. This is obtained implicitly in the proof of Proposition~2.2 in~\cite{ge2024cue}. We sketch the argument here. First we write
\begin{align*}
	X_2 &=  \sum_{|\t_j|\ge \frac{c}{N}} \left( \frac{1}{z-z_j} - \frac{1}{z_0-z_j} \right) \\
	& =  \sum_{|\t_j|\ge \frac{c}{N}}  \frac{z_0-z}{(z-z_j)(z_0-z_j)} \\
	& \ll  \frac{1}{N}\sum_{|\t_j|\ge \frac{c}{N}}  \frac{1}{|z-z_j|\cdot |z_0-z_j|}.
	\end{align*}
Now by definitions of $z$ and $z_0$ and the assumption $|\t_j| \ge c/N$ we have
	\begin{align*}
		c	|z_0-z_j| \ll  |z-z_j|.
	\end{align*}
It follows that
	\begin{align*}
X_2	& \ll \frac{1}{cN} \sum_{|\t_j|\ge \frac{c}{N}}  \frac{1}{|z_0-z_j|^2} \\
	& \ll  \frac{1}{cN} \sum_{j=1}^N  \frac{1}{|z_0-z_j|^2}.
\end{align*}
By a change of variable, transforming the unit circle into the imaginary axis, one can show that (see equation (8) in \cite{ge2024cue})
	\begin{align}
	\sum_{j=1}^N  \frac{1}{|z_0-z_j|^2} \ll N(N+|X_1|).
\end{align}
This gives
\begin{align*}
	X_2 \ll \frac{1}{c}\cdot (N+|X_1|).
\end{align*}
\qed

\begin{lem}\label{lem x3}
	We have 
\begin{align*}
	X_3   \ll  N+|X_1|   .
\end{align*}
\end{lem}
This is obtained in the proof of Proposition~2.3 in~\cite{ge2024cue}.

From the above three lemmas we can write
	\begin{align}
	\frac{P'}{P}(z) = M + E,
\end{align}
	where the main term
	\begin{align}\label{eq M}
		M = \sum_{|\t_j|<\frac{c}{N}} \frac{1}{z-z_j} 
	\end{align}
and the error term
\begin{align} \label{eq E}
	E =X_1 +X_2 -X_3 \ll \frac{1}{c} \cdot (N+|X_1|).
\end{align}

From \eqref{eq E} and \eqref{eq X1} it is clear that for every $\ell\in \mathbb Z^+$ we have
\begin{align}\label{eq E int moment}
		\mathbb E |E|^{2\ell}   \ll_\ell \left(\frac{N}{c}\right)^{2\ell} .
\end{align}
A simple application of Jensen's (or H\"older's) inequality extends this to all real moments, as follows.

\begin{prop} \label{prop E} 
	For all $K>0$ we have
\begin{align*}
		\mathbb E |E|^{K}   \ll_K \left(\frac{N}{c}\right)^K .
\end{align*}
\end{prop}

We will also prove a moment estimate for the main term $M$, as follows.

\begin{prop} \label{prop M}
	\begin{align*}
			\mathbb E |M|^{K}   	
			& \sim \dfrac{N}{2\pi} \cdot \int_{-\infty}^{\infty} \left(\dfrac{1}{\left(\frac{a}{N}\right)^2 + \t^2} \right)^{\frac{K}{2}} d\t \\
			& = \dfrac{N}{2\pi} \cdot \left(\dfrac{N}{a}\right)^{K-1} \cdot \sqrt{\pi} \cdot  \frac{ \Gamma \left(\frac{K-1}{2}\right)}{ \Gamma \left(\frac{K}{2}\right)}.
	\end{align*}
\end{prop}	

\noindent \textit{Proof of Proposition \ref{prop M}.}  We let 
\begin{align*}
	&	\mathcal T_0 =\{ U\in U(N) : U \text{ has no eigenvalues in } |\t|<c/N \}, \\
	&	\mathcal T_1 =\{ U\in U(N) : U \text{ has exactly $1$ eigenvalue in } |\t|<c/N \}, \\
	&	\mathcal T_2 =\{ U\in U(N) : U \text{ has at least $2$ eigenvalues in } |\t|<c/N \},
\end{align*}
and so, 
\begin{align*}
	\mathbb E  |M|^ = \left(\int_{\mathcal T_0} + \int_{\mathcal T_1} + \int_{\mathcal T_2} \right) \left| \sum_{|\t_j|<\frac{c}{N}} \frac{1}{z-z_j}\right|^K dU.
\end{align*}
The $\mathcal T_0$ integral is trivial $0$ since the integrand is $0$. 

Consider the $\mathcal T_1$ integral. By definition 
\begin{align*}
	\int_{\mathcal T_1} = \int_{\mathcal T_1} \left|  \frac{1}{z-e^{i\t}}\right|^K dU,
\end{align*}
where $e^{i\t}$ is the unique eigenvalue of $U$ in the region $|\t| < c/N$. We may rewrite this as 
\begin{align*}
	\int_{\mathcal T_1} = \int_{-c/N}^{c/N} \frac{1}{|z-e^{i\t}|^K}\cdot f (\t) \cdot P (\t) d \t,
\end{align*}
where 
\begin{align*}
	f(\t) = f_{U(N)} (\t) := \text{ the likelihood that } \t \text{ is an eigenangle of some } U \in U(N) 
\end{align*}
and \begin{align*}
	P(\t) := \Pr\left(\text{conditional on } \t \text{ is an eigenangle, there is exactly one eigenangle in }  \left[\dfrac{-c}{N}, \dfrac{c}{N}\right]\right).
\end{align*}
Here  the notation $\Pr(\cdot)$ means probability.
From the standard $1$-level density estimates we know that
\begin{align} \label{eq density}
	f(\t)= \frac{N}{2\pi}.
\end{align}
Next, we show 
\begin{align}\label{eq P}
	P(\t) = 1+o(1)
\end{align} 
for $\t \in [-c/N, c/N]$.
This is intuitively true since the condition~\eqref{c 1} $c=o(1)$ implies that it is not likely to see two or more eigenvalues in $[-c/N, c/N]$. To argue rigorously, we shall show that for $|\t|<c/N$ and for every positive integer $m$, 
\begin{align}\label{eq pr m}
P_m(\t) & :=	\Pr(\text{exactly } m \text{ eigenangles in } [-c/N, c/N], \text{ and one in } [\t, \t+d\t])  \notag  \\
& \le \frac{N}{2\pi} d\t \cdot m \left(\frac{c}{\pi}\right)^{m-1}.
\end{align}
To prove this, recall the $m$-level density formula for $U(N)$ is (see~\cite{conrey2005rmtntoes})
\begin{align}\label{eq m level}
&	\int_{U(N)} \sum_{\substack{J\subset \{1,...,N\}\\J=\{j_1,...,j_n\}}} F(\t_{j_1},...,\t_{j_n})  dU \notag \\[1ex]
& \quad = \frac{1}{(2\pi)^n n!}\int_{[0,2\pi]^n} F(\t_1,...,\t_n) \det _{n\times n} S_N(\t_k-\t_j) d\t_1\cdots d\t_n,
\end{align}
where \begin{align*}
	S_N(x)=\dfrac{\sin \frac{Nx}{2}}{\sin \frac{x}{2}}.
\end{align*}
Apply this with $$F= \mathbbm{1}_{[\t, \t+d\t]} \times \mathbbm{1}_{[-c/N,c/N]}^{m-1}+\mathbbm{1}_{[-c/N,c/N]}\times \mathbbm{1}_{[\t, \t+d\t]} \times \mathbbm{1}_{[-c/N,c/N]}^{m-2}+\cdots + \mathbbm{1}_{[-c/N,c/N]}^{m-1}\times \mathbbm{1}_{[\t, \t+d\t]} $$
for $|\t|<c/N$, and~\eqref{eq pr m} follows since the left-hand side of~\eqref{eq m level} is at least $P_m(\t)$ while the right-hand side of~\eqref{eq m level} is
\begin{align*}
	\le \frac{N}{2\pi} d\t \cdot m \left(\frac{c}{\pi}\right)^{m-1},
\end{align*}
where the last assertion can be seen by expanding the determinant as a permutation sum. Therefore, we have
\begin{align*}
	P_2(\t)+P_3(\t)+\cdots + P_N(\t) &  \ll Nd\t \cdot \sum_{m\ge 2} m c^{m-1} \\
	& \ll (N d\t) c ,
\end{align*}
while 
\begin{align*}
	P_1(\t) +	P_2(\t)+P_3(\t)+\cdots + P_N(\t) = f(\t) d\t = \frac{N}{2\pi} d\t.
\end{align*}
Since $c=o(1)$, it follows that 
\begin{align*}
	P_1(\t) \sim f(\t) d\t ,
\end{align*}
and thus, the conditional probability
\begin{align*}
	P(\t) = \frac{P_1(\t)  }{f(\t) d\t } \sim 1.
\end{align*}
Therefore, with $f(\t)$ and $P(\t)$ estimated, we arrive at
\begin{align*}
	\int_{\mathcal T_1}  \sim \int_{-c/N}^{c/N} \frac{1}{|z-e^{i\t}|^K}\cdot \frac{N}{2\pi} d \t. 
\end{align*}
Recall that $z=1-\frac{a}{N}$. Thus, 
\begin{align*}
	|z-e^{i\t}|^2 & =\left(1-\frac{a}{N} - \cos\t\right)^2 + \sin^2 \t \\
	& \sim \left(-\frac{a}{N} + \frac{\t^2}{2} + O(\t^4)\right)^2 +  \t^2 \\
	& \sim \left(\frac{a}{N}\right)^2 + \t^2.
\end{align*}
It follows that
\begin{align*}
	\int_{\mathcal T_1} 	 \sim \int_{-c/N}^{c/N} \left(\frac{1}{(a/N)^2+\t^2}\right)^{K/2}\cdot \frac{N}{2\pi} d \t .
\end{align*}
By a change of variable and the condition~\eqref{c 2} that $a=o(c)$, it is easy to see that
	\begin{align*}
	\int_{\mathcal T_1} \sim \int_{-\infty}^{\infty} \left(\frac{1}{(a/N)^2+\t^2}\right)^{K/2}\cdot \frac{N}{2\pi} d \t .
	\end{align*}
	It is now straightforward to compute the above integral and we conclude that
	\begin{align*}
	\int_{\mathcal T_1} \sim \dfrac{N}{2\pi} \cdot \left(\dfrac{N}{a}\right)^{K-1} \cdot \sqrt{\pi} \cdot  \frac{ \Gamma \left(\frac{K-1}{2}\right)}{ \Gamma \left(\frac{K}{2}\right)}
\end{align*}
which gives the leading term in Proposition~\ref{prop M}.

It remains to show that $$	\int_{\mathcal T_2} = o\left(	\int_{\mathcal T_1} \right).$$ 
A familiar inequality says that for $m\in \mathbb Z^+$, $s\in \mathbb R_{\ge 1}$ and $a_1, ..., a_m \in \mathbb R^+$ we have
\begin{align*}
	m\cdot 	(a_1+ \cdots + a_m)^s \le (ma_1)^s + \cdots +(ma_m)^s .
\end{align*}
This inequality holds essentially because when $s\ge1$ the function $x^s$ is convex, and one can prove it easily by first proving for $m=2$ and then applying two-term averaging repeatedly.
This inequality implies
\begin{align*}
	(a_1+ \cdots + a_m)^s \le (a_1^s + \cdots + a_m^s ) \cdot m^{s-1}.
\end{align*}
Now in the $\mathcal T_2$ integral we first apply the triangle inequality and then the above inequality to the integrand , and we obtain
\begin{align*}
	\int_{\mathcal T_2}  =	\int_{\mathcal T_2}  \left| \sum_{|\t_j|<\frac{c}{N}} \frac{1}{z-z_j}\right|^K dU  
	& \le 	\int_{\mathcal T_2}  \left( \sum_{|\t_j|<\frac{c}{N}} \frac{1}{|z-z_j|}\right)^K dU \\
	& \le   \int_{\mathcal T_2}  \left( \sum_{|\t_j|<\frac{c}{N}} \frac{1}{|z-z_j|^K}\right) m^{K-1} dU,
\end{align*}
where $m=m(U)$ is the number of $\t$'s in $[-c/N, c/N]$ for $U$. As we did in the $\mathcal T_1$ integral, we interchange summation in the last integral and  arrive that
\begin{align} \label{eq T2}
	\int_{\mathcal T_2} \le	\int_{-c/N}^{c/N} \left( \frac{1}{|z-e^{i\t}|^K} \cdot \int_{\mathcal T_2, \text{ and }\t \text{ is an eigenangle}} m^{K-1} d_\t U\right) \cdot f (\t)d \t,
\end{align}
where $d_\t U$ is the conditioning Haar measure on the subset of $U(N)$ when $\t$ is an eigenangle.  Now recall~\eqref{eq pr m} that for $\t \in [-c/N, c/N]$, 
\begin{align*}
	\Pr(\text{exactly } m \text{ eigenangles in } [-c/N, c/N], \text{ and one in } [\t, \t+d\t]) 
 \le \frac{N}{2\pi} d\t \cdot m \left(\frac{c}{\pi}\right)^{m-1}.
\end{align*}
Thus, for $\t \in [-c/N, c/N]$ we have
\begin{align*} 
& 	\Pr(\text{conditional on an eigenangle at } \t, \text{ there are exactly } m \text{ eigenangles in } [-c/N, c/N] )  \\
	& \quad  \ll m c^{m-1}.
\end{align*}
It follows that
\begin{align*}
	&	\int_{\mathcal T_2, \text{ and }\t \text{ is an eigenangle}} m^{K-1} d_\t U \\
	& \quad  = \sum_{m=2}^{N} m^{K-1} \cdot	\Pr\big(\text{conditional on an eigenangle at } \t, \\[-2.5ex]
	& \qquad \qquad \qquad \qquad \qquad \text{ there are exactly } m \text{ eigenangles in } [-c/N, c/N] \big) \\
	& \quad \ll \sum_{m=2}^{N} m^{K} \cdot c^{m-1} \\
	& \quad \ll_K c  = o_K(1),
\end{align*}
where in the last line we recall again that $c=o(1)$ by~\eqref{c 1}.
Plugging this into \eqref{eq T2} we obtain
\begin{align*}
	\int_{\mathcal T_2} & =o_K(1) \cdot \int_{-c/N}^{c/N}  \frac{1}{|z-e^{i\t}|^K} f(\t)d \t \\
	& =o_K(1) \cdot \int_{\mathcal T_1}.
\end{align*}
This finishes the proof of Proposition~\ref{prop M}. \qed

Now recall the condition~\eqref{c 3} that  $c^{-K}=o(a^{1-K})$. Thus, by Propositions~\ref{prop E} and~\ref{prop M} we see that
\begin{align}\label{eq EoM}
		\mathbb E |E|^{K}  = o\left( 	\mathbb E |M|^{K}   \right).
\end{align}
From this we shall deduce the following proposition, which together with Proposition~\ref{prop M} gives Theorem~\ref{thm u}.
\begin{prop} \label{prop ME}
	\begin{align*}
		\mathbb E |M+E|^{K}   	\sim	\mathbb E |M|^{K}   .
	\end{align*}
\end{prop}	

\noindent \textit{Proof of Proposition \ref{prop ME}.}  We write
\begin{align*}
		\mathbb E |M+E|^{K}   = \int_{|M| > 3 |E|} |M+E|^K  dU+ \int_{|M| \le 3 |E|} |M+E|^K dU.
\end{align*}
For the second integral we have
\begin{align*}
	 \int_{|M| \le 3 |E|} |M+E|^K  
	 & \le \int_{|M| \le 3 |E|} (|M|+|E|)^K dU  \\ 
	 & \le  \int_{|M| \le 3 |E|} (4|E|)^K dU \\
	 & \ll_K  \int_{U(N)} |E|^K dU \\
	 & = 	\mathbb E |E|^{K} \\
	 & = o\left( 	\mathbb E |M|^{K}   \right)
\end{align*}
by \eqref{eq EoM}. 

For the first integral we write
\begin{align*}
	A=|M|^2, \quad B=|E|^2 + M\overline{E} + \overline{M}E,
\end{align*}
and so
\begin{align*}
	 \int_{|M| > 3 |E|} |M+E|^K  dU & =  \int_{|M| > 3 |E|} (|M|^2 + |E|^2 + M\overline{E} + \overline{M}E) ^{K/2}  dU \\
	 & =  \int_{|M| > 3 |E|} (A + B) ^{K/2}  dU \\
	 & = \int_{|M| > 3 |E|}  |M|^K \left(1+\frac{B}{A}\right)^{K/2} dU \\
	 & = \int_{|M| > 3 |E|}  |M|^K \left(1+\frac{K}{2}\frac{B}{A} + O_K\left(\frac{B^2}{A^2}\right)\right) dU,
\end{align*}
where in the last line we have used the Taylor expansion with remainder that for real $\alpha$ and $|x|<\frac{7}{9}$, say, 
\begin{align*}
	(1+x)^{\alpha} = 1+ \alpha x +O_\alpha (x^2),
\end{align*}
and note that the condition $|M| > 3 |E|$ guarantees $|B/A| < 7/9$. Thus, 
\begin{align*}
	\int_{|M| > 3 |E|} |M+E|^K  dU & = \int_{|M| > 3 |E|}  |M|^K dU + O_K \left( \int |M|^{K-1}|E| + |M|^{K-2}|E|^2 dU\right).
\end{align*}
As before we have 
\begin{align*}
	 \int_{|M| \le 3 |E|} |M|^K   = o\left( 	\mathbb E |M|^{K}   \right), 
\end{align*}
so that
\begin{align*}
	\int_{|M| > 3 |E|}  |M|^K dU \sim \mathbb E  |M|^K.
\end{align*}
For the $O_K$-terms we use H\"older's inequality together with Propositions~\ref{prop E} and~\ref{prop M} to obtain again an upper bound $  o\left( 	\mathbb E |M|^{K}   \right) $.  This finishes the proof of Proposition~\ref{prop ME}. \qed

\section{Orthogonal and Symplectic ensembles}

In this section we give a proof sketch for Theorems~\ref{thm so} and~\ref{thm usp}.  We shall only indicate the differences from the unitary case.

The three lemmas cited from~\cite{ge2024cue} can be used or easily adapted.  For example, Lemma~\ref{lem x1} (that is, even integer moments bounds for $X_1$) is proved by applying a ratios theorem in~\cite{conrey2008correlations} together with a discretization argument. The analogous results for orthogonal and symplectic ensembles can be deduced by applying the ratios theorems in~\cite{snaith2018orthogonal}  for these ensembles together with a similar discretization argument as in~\cite{ge2024cue}.

In $SO(2N)$, eigenvalues appear in conjugate pairs, and we shall only label eigenvalues in the upper half plane, as $z_1, ..., z_N$. This explains why the integral in~\eqref{thm eq SO} is from $0$ rather than $-\infty$. 
Also, in the corresponding summation~\eqref{eq M} for $M$ we group conjugate pairs and so each summand looks like
\begin{align*}
	\frac{1}{z-z_j} + 	\frac{1}{z-\overline{z_j}} = 2 \Re 	\frac{1}{z-z_j} ,
\end{align*}
and if $\t_j < c/N$ the above is
 \begin{align*}
	\sim \dfrac{2\cdot {\left( - \frac{a}{N} + O(\t^2)\right) }}{\left(\frac{a}{N}\right)^2 + \t_j^2} .
\end{align*}
The $O$-term makes a small contribution toward the integral, and this explains the shape of the integrand in the integral in~\eqref{thm eq SO}. The likelihood function $f_{SO(2N)}$ can be calculated using $1$-level density estimate for $SO(2N)$ and the result is
\begin{align*}
	f_{SO(2N)} \sim \frac{2N}{\pi},
\end{align*}
and this should replace~\eqref{eq density}. The $P(\t)$ and $P_m(\t)$ terms in~\eqref{eq P} and~\eqref{eq pr m} can be estimated in the same way and the results are similar as the unitary case. The main term in~\eqref{thm eq SO} is of order $N^K a^{1-K}$, and so we require $K>1$ in order for the parameter $c$ to exist.

In $USp(2N)$, eigenvalues again appear in conjugate pairs, and so the integral in~\eqref{thm eq USp} is from $0$.
As in $SO(2N)$, summands in the corresponding summation~\eqref{eq M} for $M$ are
 \begin{align*}
	\sim \dfrac{2\cdot {\left( - \frac{a}{N} + O(\t^2)\right) }}{\left(\frac{a}{N}\right)^2 + \t_j^2} .
\end{align*}
 if $\t_j < c/N$, and again the $O$-term is negligible since its contribution toward the integral is small. The likelihood function $f_{USp(2N)}$ can be calculated using $1$-level density estimate for $USp(2N)$ and the result is
\begin{align*}
	f_{USp(2N)} \sim \frac{2 N^3 \t^2}{ 3 \pi}.
\end{align*}
The $P_m(\t)$ terms can be treated in a similar way (with a bit more care dealing with Taylor expansions) and the bounds look like
\begin{align*}
	N^3\t^2 d\t \cdot m (c^3)^{m-1}.
\end{align*}
From this we again have $P(\t) \sim 1$.
This time the main term in~\eqref{thm eq USp} is of order $N^K a^{3-K}$, so we require $K>3$ to guarantee the existence of $c$.

\section{Acknowledgment}

The author would like to thank Nina Snaith for very helpful communication.


\newcommand{\etalchar}[1]{$^{#1}$}


\begin{thebibliography}{BBB{\etalchar{+}}19}
	
	\bibitem[ABS23]{alvarez2023noninteger}
	Emilia Alvarez, Pierre Bousseyroux, and Nina~C. Snaith.
	\newblock Asymptotics of non-integer moments of the logarithmic derivative of
	characteristic polynomials over {$SO(2N+1)$}.
	\newblock {\em arXiv preprint arXiv:2303.07813}, 2023.
	
	\bibitem[AS20]{alvarez2020moments}
	E.~Alvarez and N.~C. Snaith.
	\newblock Moments of the logarithmic derivative of characteristic polynomials
	from {$SO(N)$} and {$USp(2N)$}.
	\newblock {\em J. Math. Phys.}, 61(10):103506, 32, 2020.
	
	\bibitem[Bal16]{baluyot2016ah}
	Siegfred Alan~C. Baluyot.
	\newblock On the pair correlation conjecture and the alternative hypothesis.
	\newblock {\em J. Number Theory}, 169:183--226, 2016.
	
	\bibitem[BBB{\etalchar{+}}19]{bailey2019mixedmoments}
	E.~C. Bailey, S.~Bettin, G.~Blower, J.~B. Conrey, A.~Prokhorov, M.~O.
	Rubinstein, and N.~C. Snaith.
	\newblock Mixed moments of characteristic polynomials of random unitary
	matrices.
	\newblock {\em J. Math. Phys.}, 60(8):083509, 26, 2019.
	
	\bibitem[CFZ08]{conrey2008autocorrelationofratios}
	Brian Conrey, David~W. Farmer, and Martin~R. Zirnbauer.
	\newblock Autocorrelation of ratios of {$L$}-functions.
	\newblock {\em Commun. Number Theory Phys.}, 2(3):593--636, 2008.
	
	\bibitem[Con05]{conrey2005rmtntoes}
	Brian Conrey.
	\newblock Notes on eigenvalue distributions for the classical compact groups.
	\newblock In {\em Recent perspectives in random matrix theory and number
		theory}, volume 322 of {\em London Math. Soc. Lecture Note Ser.}, pages
	111--145. Cambridge Univ. Press, Cambridge, 2005.
	
	\bibitem[CS08]{conrey2008correlations}
	John~Brian Conrey and Nina~Claire Snaith.
	\newblock Correlations of eigenvalues and {R}iemann zeros.
	\newblock {\em Commun. Number Theory Phys.}, 2(3):477--536, 2008.
	
	\bibitem[FGLL13]{farmer2013logderiv}
	D.~W. Farmer, S.~M. Gonek, Y.~Lee, and S.~J. Lester.
	\newblock Mean values of {$\zeta'/\zeta(s)$}, correlations of zeros and the
	distribution of almost primes.
	\newblock {\em Q. J. Math.}, 64(4):1057--1089, 2013.
	
	\bibitem[FK12]{farmer2012landau}
	David~W. Farmer and Haseo Ki.
	\newblock Landau-{S}iegel zeros and zeros of the derivative of the {R}iemann
	zeta function.
	\newblock {\em Adv. Math.}, 230(4-6):2048--2064, 2012.
	
	\bibitem[Ge17a]{ge2017gaps}
	Fan Ge.
	\newblock The distribution of zeros of {$\zeta'(s)$} and gaps between zeros of
	{$\zeta(s)$}.
	\newblock {\em Adv. Math.}, 320:574--594, 2017.
	
	\bibitem[Ge17b]{ge2017number}
	Fan Ge.
	\newblock The number of zeros of {$\zeta'(s)$}.
	\newblock {\em Int. Math. Res. Not. IMRN}, (5):1578--1588, 2017.
	
	\bibitem[Ge23]{ge2023zeta}
	Fan Ge.
	\newblock Mean values of the logarithmic derivative of the riemann
	zeta-function near the critical line.
	\newblock {\em Mathematika}, 69(2):562--572, 2023.
	
	\bibitem[Ge24]{ge2024cue}
	Fan Ge.
	\newblock On the logarithmic derivative of characteristic polynomials for random unitary matrices.
	\newblock {\em Bull. Lond. Math. Soc.}, 56(3):1114--1128, 2024.
	
	\bibitem[GGM01]{goldston2001pair}
	D.~A. Goldston, S.~M. Gonek, and H.~L. Montgomery.
	\newblock Mean values of the logarithmic derivative of the {R}iemann
	zeta-function with applications to primes in short intervals.
	\newblock {\em J. Reine Angew. Math.}, 537:105--126, 2001.
	
	\bibitem[GM87]{goldston1987pair}
	Daniel~A. Goldston and Hugh~L. Montgomery.
	\newblock Pair correlation of zeros and primes in short intervals.
	\newblock In {\em Analytic number theory and {D}iophantine problems
		({S}tillwater, {OK}, 1984)}, volume~70 of {\em Progr. Math.}, pages 183--203.
	Birkh\"{a}user Boston, Boston, MA, 1987.
	
	\bibitem[Gol88]{goldston1988pair}
	D.~A. Goldston.
	\newblock On the pair correlation conjecture for zeros of the {R}iemann
	zeta-function.
	\newblock {\em J. Reine Angew. Math.}, 385:24--40, 1988.
	
	\bibitem[Guo96]{guo1996zeros}
	Charng~Rang Guo.
	\newblock On the zeros of the derivative of the {R}iemann zeta function.
	\newblock {\em Proc. London Math. Soc. (3)}, 72(1):28--62, 1996.
	
	\bibitem[GY07]{garaev2007small}
	M.Z. Garaev and C.Y. Y{\i}ld{\i}r{\i}m.
	\newblock On small distances between ordinates of zeros of {$\zeta(s)$} and
	{$\zeta'(s)$}.
	\newblock {\em Int. Math. Res. Not. IMRN}, (21):Art. ID rnm091, 14, 2007.
	
	\bibitem[Ki08]{ki2008zeros}
	Haseo Ki.
	\newblock The zeros of the derivative of the {R}iemann zeta function near the
	critical line.
	\newblock {\em Int. Math. Res. Not. IMRN}, (16):Art. ID rnn064, 23, 2008.
	
	\bibitem[KS99]{katz1999random}
	Nicholas~M. Katz and Peter Sarnak.
	\newblock {\em Random matrices, {F}robenius eigenvalues, and monodromy},
	volume~45 of {\em American Mathematical Society Colloquium Publications}.
	\newblock American Mathematical Society, Providence, RI, 1999.
	
	\bibitem[KS00]{keating2000momentszeta}
	J.~P. Keating and N.~C. Snaith.
	\newblock Random matrix theory and {$\zeta(1/2+it)$}.
	\newblock {\em Comm. Math. Phys.}, 214(1):57--89, 2000.
	
	\bibitem[Les14]{lester2014distribution}
	S.~J. Lester.
	\newblock The distribution of the logarithmic derivative of the {R}iemann
	zeta-function.
	\newblock {\em Q. J. Math.}, 65(4):1319--1344, 2014.
	
	\bibitem[LM74]{levinson1974derivatives}
	Norman Levinson and Hugh~L. Montgomery.
	\newblock Zeros of the derivatives of the {R}iemann zeta-function.
	\newblock {\em Acta Math.}, 133:49--65, 1974.
	
	\bibitem[Mon73]{montgomery1973pair}
	H.~L. Montgomery.
	\newblock The pair correlation of zeros of the zeta function.
	\newblock In {\em Analytic number theory ({P}roc. {S}ympos. {P}ure {M}ath.,
		{V}ol. {XXIV}, {S}t. {L}ouis {U}niv., {S}t. {L}ouis, {M}o., 1972)}, pages
	181--193, 1973.
	
	\bibitem[MS18]{snaith2018orthogonal}
	A.~M. Mason and N.~C. Snaith.
	\newblock Orthogonal and symplectic {$n$}-level densities.
	\newblock {\em Mem. Amer. Math. Soc.}, 251(1194):v+93, 2018.
	
	\bibitem[MV74]{mont1974hilbert}
	H.~L. Montgomery and R.~C. Vaughan.
	\newblock Hilbert's inequality.
	\newblock {\em J. London Math. Soc. (2)}, 8:73--82, 1974.
	
	\bibitem[Rad14]{radziwill2014gaps}
	Maksym Radziwi\l\l.
	\newblock Gaps between zeros of {$\zeta(s)$} and the distribution of zeros of
	{$\zeta'(s)$}.
	\newblock {\em Adv. Math.}, 257:6--24, 2014.
	
	\bibitem[Sel43]{selberg1943prime}
	Atle Selberg.
	\newblock On the normal density of primes in small intervals, and the
	difference between consecutive primes.
	\newblock {\em Arch. Math. Naturvid.}, 47(6):87--105, 1943.
	
	\bibitem[Sel46]{selberg1946contributions}
	Atle Selberg.
	\newblock Contributions to the theory of the {R}iemann zeta-function.
	\newblock {\em Arch. Math. Naturvid.}, 48(5):89--155, 1946.
	
	\bibitem[Sou98]{soundararajan1998horizontal}
	K.~Soundararajan.
	\newblock The horizontal distribution of zeros of {$\zeta'(s)$}.
	\newblock {\em Duke Math. J.}, 91(1):33--59, 1998.
	
	\bibitem[Zha01]{zhang2001zeros}
	Yitang Zhang.
	\newblock On the zeros of {$\zeta'(s)$} near the critical line.
	\newblock {\em Duke Math. J.}, 110(3):555--572, 2001.
	
\end{thebibliography}
\end{document}